\documentclass[aps,prl,showpacs,twocolumn,superscriptaddress,amsmath,amssymb]{revtex4}
\usepackage{graphicx}% Include figure files
\usepackage{amsmath}
\usepackage{amssymb}
\usepackage{enumerate}
\usepackage{tikz}
\usepackage{pgf}
\usepackage{color}
\begin{document}
\title{Neutral gas sympathetic cooling of an ion in a Paul trap}
\affiliation{Department of Physics and Astronomy, University of California, Los Angeles, California 90095, US}
\author{Kuang Chen, Scott T. Sullivan, and Eric R. Hudson}
\begin{abstract}
A single ion immersed in a neutral buffer gas is studied. An analytical model is developed that gives a complete description of the dynamics and steady-state properties of the ions. An extension of this model, using techniques borrowed from the mathematics of finance, is used to explain the recent observation of non-Maxwellian statistics for these systems. Taken together, these results offer an explanation of the longstanding issues associated with sympathetic cooling of an ion by a neutral buffer gas. 
\end{abstract}

\maketitle

The fact that two isolated objects in thermal contact tend to the same temperature is the most basic tenet of thermodynamics. It is also the essence of the technique of sympathetic cooling, where a sample is prepared at a desired temperature by bringing it into thermal contact with a much larger body already at the desired temperature. It is difficult to overstate the importance of this technique as it underpins applications ranging from basic refrigeration to quantum information science. 

It may be considered surprising then that a gas of ions trapped in a radio-frequency Paul trap and immersed in a reservoir of neutral atoms, does not equilibrate to the same temperature as the neutral atoms. Instead, the ions are found to have a higher temperature than the neutral gas, and in some cases are heated so much that they escape the trap. Since the early work of Major and Dehmelt \cite{Major1968} it has been known that this apparent contradiction with the laws of thermodynamics is due to the fact that ions are subject to a time-dependent confining potential and are therefore not an isolated system. However, despite pioneering work by Dehmelt and others \cite{Moriwaki1992, Vedel1983}, an accurate analytical description of the relaxation process has not yet been achieved. Given the recent surge in interest in hybrid atom-ion systems \cite{Grier2009, Zipkes2010a, Zipkes2010b, Hall2011, Rellergert2011, Rellergert2013, Sullivan2012, Schmid2010a, Ratschbacher2012, Ratschbacher2013}, where ions are immersed in baths of ultracold atoms, there is currently a strong need for such a description so that these systems can be understood and optimized.  

Building upon the important work of Moriwaki {\textit {et al}}. \cite{Moriwaki1992}, here we present a simple kinematic model, which accurately describes the ion relaxation process. This model, which has been verified by detailed molecular dynamics simulations, provides a simple and accurate means to calculate both the relaxation dynamics and the properties of the ion steady state. This model also provides significant physical intuition for the problem and as such suggests several ways for optimizing ongoing and planned experiments in fields as diverse as quantum chemistry \cite{Grier2009, Zipkes2010a, Zipkes2010b, Hall2011, Rellergert2011, Rellergert2013, Sullivan2012, Schmid2010a, Ratschbacher2012, Ratschbacher2013}, mass spectrometry \cite{Drewsen2004}, and quantum information \cite{Hudson2009}.

In the remainder of this work, we first review the basics of ion trapping and introduce the time-averaged ion kinetic energy. We then consider the effect of a collision with a neutral particle on the evolution of the kinetic energy of a single ion in a Paul trap and show that due to the presence of the time-dependent potential the collision center-of-mass frame energy is not conserved. Following this result, we develop a rate equation model, which accounts for the relaxation and exchange of the ion energy in all three dimensions. We then present simple formulae for the calculation of the ion temperature relaxation rate and steady-state value, as well as the dependency of these values on the ion trapping parameters and particle masses. We establish the validity of these results by comparing them to a detailed molecular dynamics simulation. We conclude with an explanation for the recent observation \cite{DeVoe2009} of non-Maxwellian distribution functions for these systems.

\textit{Ion trap dynamics} --
The trajectory, $r_j$, and velocity, $v_j$,  of an ion in a linear Paul trap can be expanded as a linear superposition of two orthogonal Mathieu functions $c(a_j,q_j;\tau)$ and $s(a_j,q_j;\tau)$ with coefficients $A_j$ and $B_j$,  
\begin{equation}
  \begin{aligned}
    \label{eq:ab_to_xv}
    r_j(\tau) &= A_j~c_j(\tau) + B_j~ s_j(\tau) \\
    v_j(\tau) &= A_j ~ \dot{c}_j(\tau) + B_j ~ \dot{s}_j(\tau)
  \end{aligned}
\end{equation}
where $j =x,y,z$ and the dependence on the Mathieu parameters ($\{ a_x, a_y, a_z\}  = \{-a, -a, 2a\}$ and $\{q_x, q_y,q_z\}  = \{q, -q, 0\}$ with $q=\frac{4eV_{rf}}{mr_0^2\Omega^2}$ and $a=\frac{4\alpha eU_{ec}}{mz_0^2\Omega^2}$) is suppressed~\cite{Major1968}. The Fourier transform of $c_j(\tau)$ and $s_j(\tau)$ is a discrete spectrum, 
\begin{equation}
c_j(\tau) + \imath s_j(\tau) = \sum_{n=-\infty}^{\infty} C_{2n} e^{\imath(\beta_j + 2n)\tau}.
\label{eq:Mathieu_fns}
\end{equation}
The $n=0$ term corresponds to the `typical' motion of a harmonic oscillator -- i.e. the secular ion motion. The remaining terms with $n\neq 0$ represent the components of the ion motion driven by the rf field -- i.e. the so-called micromotion. 

As a result of this spectrum, the instantaneous kinetic energy is not a conserved quantity. Instead, energy coherently flows back and forth between the kinetic energy of the ion and the confining electric field at frequency $\Omega$. Therefore, it is useful to define the time-averaged kinetic energy
\begin{equation}
  \label{eq:w_j} 
  W_j=  \frac{m}{2}\lim_{T\rightarrow\infty}\frac{1}{2T}\int_{-T}^T  v_j^2 d\tau =\frac{m}{2}\overline{\dot{c}_j^2}(A_j^2+B_j^2),
\end{equation}
where the bar denotes the time average. $W_j$ includes contributions from both the random thermal motion of the ion, i.e. the secular energy, and the micromotion. The ratio of the secular energy, $U_j$, to the total average kinetic energy is simply
\begin{equation}
  \eta_j \equiv \frac{U_j}{W_j} = \frac{|C_0|^2}{\sum_{n=-\infty}^{\infty} |C_{2n}|^2}.
\end{equation}
In the $x$ and $y$ directions, $\eta_{x,y} \approx \frac{1}{2}$ for $q<0.4$ and the micromotion energy is given by $W_{mm,j} = W_j-U_j$. In the $z$ direction where the trapping field is time-independent ($q=0$), $c_z(\tau)$ and $s_z(\tau)$ simply become the cosine and sine functions. Thus, all micromotion sidebands vanish and $\eta_z = 1$. 

\textit{Modeling the collision process} -- When a trapped ion is immersed in a buffer gas of neutral atoms, the Mathieu trajectory of the ion is modified by interactions with the neutral atoms. The ion-neutral interaction potential is comprised of a long-range attraction $V(r) = -C_4/2r^4$ and short-range repulsion, where $C_4$ is given by $C_4 = \alpha e^2/(4\pi\epsilon_0)^2$, and $\alpha$ is the polarizability of the neutral atom. Recent work \cite{Cetina2012}, has explored effects of this potential at ultracold temperatures, showing that the perturbations of the ion trajectory by the $C_4$ potential can lead to heating of the ion. Here we do not consider this effect, but given that the characteristic length of the $C_4$ interaction \cite{Gao2010} is small compared to the trap dimension we treat the collision as a point-like interaction. As will be seen, this approximation is justified, despite the important result of Ref. \cite{Cetina2012}, as the effects considered here typically lead to temperatures that preclude the observation of the effects considered in Ref. \cite{Cetina2012}.  We also make the additional simplifying assumptions that the density of the neutral atoms is constant and that inelastic processes, such as charge exchange, do not occur. 

Because the motion of the ion differs significantly in the radial and axial directions of a linear Paul trap, the relaxation and redistribution of energy is significantly more complicated than in a time-independent harmonic trap~\cite{DeCarvalho1999}. We therefore describe the statistically-averaged evolution of ion kinetic energy $\mathbf{W} = [W_x, W_y, W_z]^{\mathrm{T}}$ by a three-dimensional rate equation,
  \begin{equation}
    \label{eq:energy_relaxation_eq}
    \frac{\text{d}\langle\mathbf{W}(t)\rangle}{\text{d}t} = - \Gamma \mathbf{M} (\langle\mathbf{W}(t) \rangle - \mathbf{W}_{st}) 
  \end{equation}
where $\Gamma$ is an average collision rate (which may depend on energy), $\mathbf{M}$ is a 3$\times$3 ``relaxation matrix'' that accounts for energy damping and redistribution among the three trap directions, and $\mathbf{W}_{st}$ is the steady-state kinetic energy. The angled bracket denotes the statistical average after the sympathetic cooling experiment is repeated multiple times.

In order to calculate both $\Gamma$ and $\mathbf{M}$ it is necessary to know the neutral-ion differential elastic scattering cross-section $\text{d}\sigma_{el}/\text{d}\Omega$, which, given an interaction potential, is a straightforward quantum scattering calculation~\cite{Friedrich2005}. Regardless of the specific atom-ion potential, however, several generic arguments can be made. First, the differential cross-section always exhibits a large forward scattering peak at all energy scales~\cite{Zhang2009}. Thus, the majority of atom-ion collisions lead to only slightly deflected trajectories, resulting in a very small change in $\mathbf{W}$. Therefore, as originally argued by Dalgarno and co-workers~\cite{Dalgarno1958},  to prevent an overestimate of the energy redistribution due to collisions the momentum transfer (diffusion) differential cross-section, i.e. $\frac{d\sigma_d}{d\Omega}=\frac{d\sigma_{el}}{d\Omega}(1-\cos\theta)$ should be used to calculate the total atom-ion collision rate. Second (and fortuitously), the diffusion differential cross-section is approximately isotropic in scattering angle, especially after thermal averaging, and agrees quite well with the simple Langevin cross-section~\cite{Langevin1905} $\sigma_d \approx \sigma_L = \pi\sqrt{\frac{2C_4}{E}}$ -- see Appendix A for a comparison of a quantum scattering calculation to the Langevin differential cross section. Therefore, we replace the cross-section by an isotropic profile which integrates to $\sigma_L$. Under this approximation, the average collision rate $\Gamma = 2\pi\rho\sqrt{\frac{C_4}{\mu}}$ becomes energy independent and the calculation of $\mathbf{M}$ is greatly simplified. As demonstrated below, the validity of this approximation is confirmed by comparison to a detailed molecular dynamics simulation, which uses the full quantum differential cross-section. The resulting error in the relaxation rate is smaller than 25\% for collision energies down to 1~mK.

With the collision rate in hand, the relaxation matrix $\mathbf{M}$ is calculated by considering the kinematics of a collision between an ion and neutral atom as follows. Suppose that at time $\tau_c$ an ion undergoes an elastic collision with an incoming neutral atom of mass $m_n$ and velocity $\mathbf{v}_n$. Conservation of momentum and energy for the collision dictates that the velocity of the ion after the collision with neutral atom is given by the sum of center-of-mass velocity and the scattered relative velocity \cite{Zipkes2011}, 
\begin{equation}
  \label{eq:post_collision_velocity}
  \mathbf{v}' = \frac{1}{1+\tilde{m}}\mathbf{v} + \frac{\tilde{m}}{1+\tilde{m}}\mathbf{v}_n + \frac{\tilde{m}}{1+\tilde{m}}\mathcal{R}(\mathbf{v} - \mathbf{v}_n)
\end{equation}
where $\tilde{m} = \frac{m_n}{m_i}$ is the mass ratio and $\mathcal{R}$ is the collision rotation matrix, which following the above discussion is isotropic. Likewise, because the characteristic length of the $C_4$ interaction \cite{Gao2010} is small compared to the trap dimension, the position of the ion is assumed to be unchanged during the collision, i.e. $\mathbf{r}'=\mathbf{r}$. By requiring that $\mathbf{r}'$ and $\mathbf{v}'$ also correspond to a Mathieu solution through Eq. \ref{eq:ab_to_xv}, a new set of oscillation amplitude $(A_j', B_j')$ and thus, the average kinetic energy after the collision $\mathbf{W}'$ can be found. 

This last step is the critical difference between sympathetic cooling in static and time-dependent traps, which is illustrated with the following one-dimensional example. In a static trap, like that in Ref. \cite{Campbell2007}, if a collision happens at position $x = a$ that reduces the velocity such that $v_x'=0$, a trapped particle of mass $m$ begins a `new' oscillation trajectory, $x' = a \cos(2\pi\sqrt{k/m} ~t)$,  where $k$ is the trap spring constant. This collision always reduces the total energy of the particle. By contrast in the time-dependent potential of a linear Paul trap, because of the terms in Eq.~\ref{eq:Mathieu_fns} with $n \neq 0$, it is possible that even though the collision brings the particle to rest, the particle may have a \textit{higher} energy after the collision. 

This can be seen by again considering a collision that leads to $v_x'=0$, which depending on the rf phase could be accomplished by having large and opposite contributions to the velocity from the $n = 0$ (secular) mode and $n\neq 0$ (micromotion) modes.  Thus, even though the particle is momentarily stopped, it could leave the collision on a trajectory of higher amplitude.

With this prescription the calculation of $\mathbf{M}$ is straightforward and proceeds as follows (see Appendix B for full details). First we rewrite Eq.~\ref{eq:w_j} in terms of the instantaneous coordinates for the $x$ direction and find the change in $W_{x}$ per collision as:
 \begin{equation}
  \begin{aligned}
    \label{eq:delta_wx}
     W_x' -  W_x & =   - \frac{m\overline{\dot{c}_x^2}}{w_{0x}^2}(c_x\dot{c}_x+s_x\dot{s}_x)(x( v_x' -v_x))\\
     & \;\;\;\; + \frac{m \overline{\dot{c}_x^2}}{2w_{0x}^2} (c_x^2+s_x^2)( v_x'^2-v_x^2) \\
     &\equiv \Delta W_{x,1} +  \Delta W_{x,2} . 
  \end{aligned}
\end{equation}
Then we take the statistical average of Eq. \ref{eq:delta_wx} over $\mathbf{v_n}, \mathcal{R}$ and collision time $\tau_c$. Since both $\langle \mathbf{v_n}\rangle$ and $\langle \mathcal{R}(\mathbf{v}-\mathbf{v_n}) \rangle $ vanish, $\langle v_{x}' \rangle = \frac{1}{1+\tilde{m}}v_x$, and $\langle \Delta W_{x,1} \rangle  = \frac{\tilde{m}}{1+\tilde{m}}\epsilon_x \langle W_x \rangle$, where $ \epsilon_x  = \frac{\overline{(c_x\dot{c}_x+s_x\dot{s}_x)^2}}{w_{0x}^2}$. Likewise, noting that since $\mathbf{v_n}$, $\mathbf{v}$ and $\mathcal{R}$ are uncorrelated the average value of cross-correlation terms between them vanish and that $\mathcal{R}$ is random rotation, $\langle[\mathcal{R}(\mathbf{v}-\mathbf{v_n})]_x^2\rangle = \frac{1}{3}\langle(\mathbf{v}-\mathbf{v_n})^2\rangle$, we have
\begin{equation}
  \begin{aligned}
  \label{eq:delta_wx_2}
  \langle \Delta W_{x,2} \rangle = \frac{\tilde{m}}{(1+\tilde{m})^2}\Big( \Big(-\frac{2\tilde{m}+2}{3}\Big)(1+\epsilon_x) \langle W_x \rangle \\+\frac{\tilde{m}\alpha_x}{6} \langle W_y \rangle +\frac{\tilde{m}\alpha_x}{6} \langle W_z \rangle + \alpha_x \langle W_n \rangle \Big),\nonumber
\end{aligned}
\end{equation}
where $\alpha_x  = \frac{\overline{(c_x^2+s_x^2)}\cdot\overline{(\dot{c}_x^2+\dot{s}_x^2)}}{w_{0,x}^2}$, and $\langle W_n \rangle$ is the average kinetic energy of neutral atom in each direction.

Combining the results of $\langle \Delta W_{x,1} \rangle $ and $\langle \Delta W_{x,2} \rangle $, and the results for the $y$ and $z$ directions, finally we have 
\begin{align}
    \langle\mathbf{W'}\rangle - \langle\mathbf{W}\rangle & = -\mathbf{M}\langle\mathbf{W}\rangle + \mathbf{N} \nonumber \\
    & = -\mathbf{M}(\langle\mathbf{W}\rangle-\mathbf{W}_{st})
\end{align}
where  
  \begin{equation}
    \label{eq:M}
    \mathbf{M} = - \frac{\tilde{m}^2}{(1+\tilde{m})^2}
    \begin{bmatrix}
      \frac{2\epsilon-1}{3}-\frac{1}{\tilde{m}} & \frac{\alpha}{6} & \frac{\alpha}{6} \\
      \frac{\alpha}{6}  &           \frac{2\epsilon-1}{3}-\frac{1}{\tilde{m}}  & \frac{\alpha}{6} \\
      \frac{1}{6} &     \frac{1}{6} &  -\frac{1}{3}-\frac{1}{\tilde{m}}      \\
    \end{bmatrix}
  \end{equation}
and
  \begin{equation}
    \label{eq:N}
    \mathbf{N} = \frac{\tilde{m}}{(1+\tilde{m})^2}
    \begin{bmatrix}
      \alpha \langle W_n \rangle \\
      \alpha \langle W_n \rangle \\
      \langle W_n \rangle
    \end{bmatrix}.
  \end{equation}
And, the components of the steady-state kinetic energy $\mathbf{W}_{st} = -\mathbf{M}^{-1}\mathbf{N}$ reduce to,
\begin{equation}
  \begin{aligned}
  \label{eq:Wst_simplified}
    \frac{W_{st,x}}{\langle W_n \rangle} = \frac{W_{st,y}}{\langle W_n \rangle} &= \frac{9(2+\tilde{m})\alpha}{18-3\tilde{m}(\alpha+4\epsilon-4)-2\tilde{m}^2(\alpha+2\epsilon-1)} \\
    \frac{W_{st,z}}{\langle W_n \rangle} & = \frac{3(6+\tilde{m}(2+\alpha-4\epsilon))}{18-3\tilde{m}(\alpha+4\epsilon-4)-2\tilde{m}^2(\alpha+2\epsilon-1)} 
  \end{aligned}
\end{equation}
where $\alpha \equiv \alpha_x = \alpha_y$ and $\epsilon \equiv \epsilon_x = \epsilon_y$. Because in the $z$ direction the trapping field is time-independent, $\alpha_z=1$ and $\epsilon_z=0$.
For low values of $q$ and $a$, the numerical values of $\alpha$ and $\epsilon$ are approximated by \cite{Chen2013},
\begin{align}
  \alpha \approx 2+2q^{2.24}\\
  \epsilon \approx 1+2.4q^{2.4}
\end{align}

\begin{figure}[!]
  \centering
  \includegraphics[width=\columnwidth]{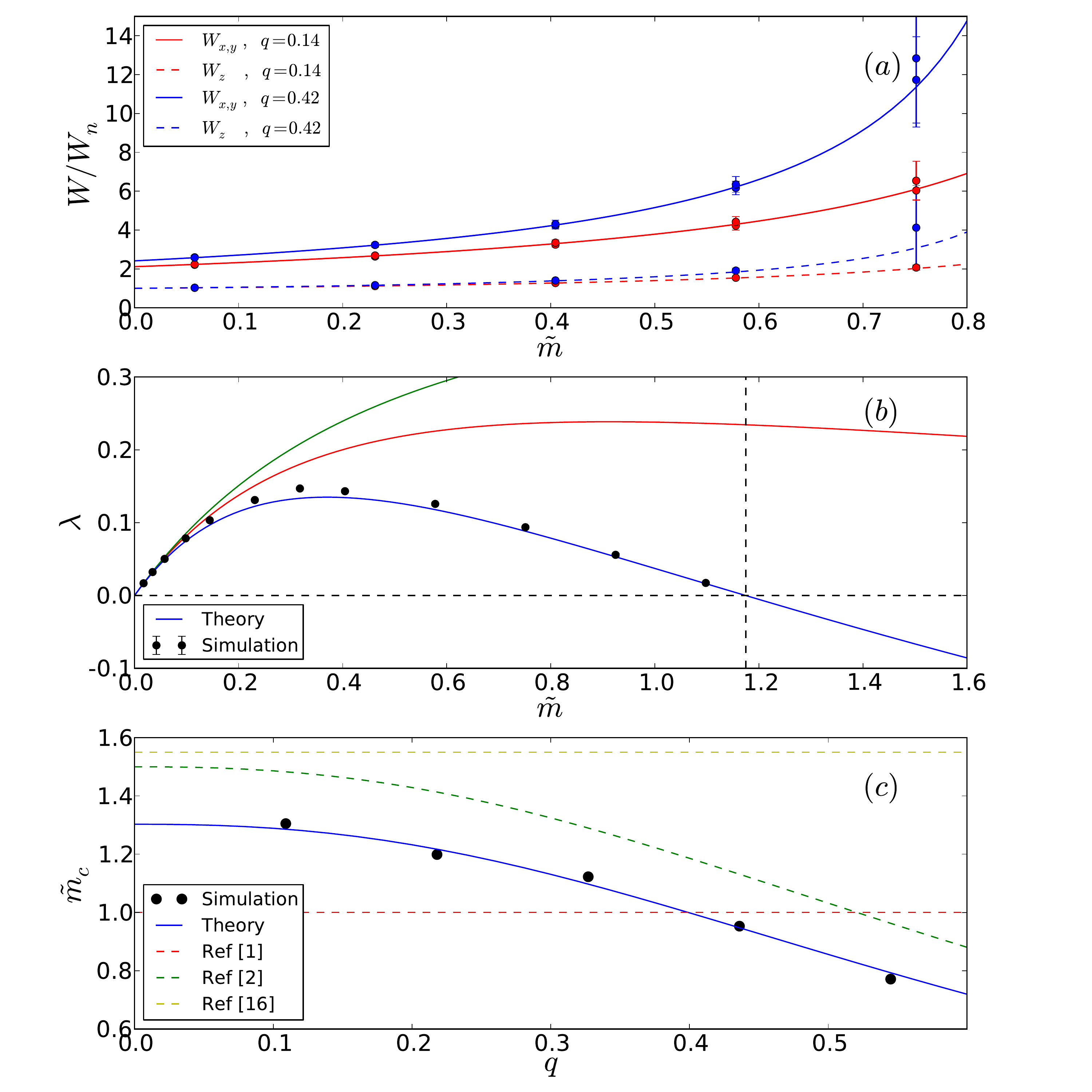}
  %\resizebox{1\columnwidth}{!}{\input{fig1.pdf}}
    \caption{(a) $\mathbf{W}_{st}$ as a function of $\tilde{m}$ for $q = 0.14$ (red) and $q=0.42$ (blue). The axial and radial components of $\mathbf{W}_{st}$  are denoted by dashed and solid lines (theory) and dots (simulation). (b) Eigenvalues of $\mathbf{M}$ as a function of $\tilde{m}$ for fixed $q=0.14$ and $a=0$. Black dots are asymptotic relaxation rates (normalized by $\Gamma$) from numerical simulations. Lines are three calculated eigenvalues of $\mathbf{M}$. The smallest one (blue line) intersects $\lambda=0$ line at $\tilde{m}=\tilde{m}_c$, which separates cooling from heating. (c) Simulated (dots) and calculated (blue line) critical mass ratio $\tilde{m}_c$ as a function of trap $q$ parameter, as compared to previous results in Ref. \cite{Moriwaki1992, Major1968, DeVoe2009}. } 
    \label{fig:ke_comparison}
\end{figure}

\textit{Model results} --
First, shown in Fig.~\ref{fig:ke_comparison}(a) are the components of $\mathbf{W}_{st}$ normalized by $\langle W_n \rangle$ obtained  from Eq. \ref{eq:Wst_simplified}. Also, shown in this figure are the results of a detailed molecular dynamics simulation, described in Appendix C. In the limit of a light neutral atom ($\tilde{m}\approx 0$) and $q\rightarrow0$, $\alpha\approx 2$, $\mathbf{W}_{st} / \langle W_n \rangle \approx [2, 2, 1]^\mathrm{T}$. Thus, at steady state,
\begin{equation}
  \label{eq:equi-partition}
  \langle U_x \rangle = \langle U_y \rangle = \langle U_z \rangle = \langle W_{mm,x} \rangle = \langle W_{mm,y} \rangle = \langle W_n \rangle ,
\end{equation}
a result often referred to as the ``equipartition'' \cite{Baba2002} of kinetic energy between secular motion and micro-motion. As $\tilde{m}$ increases, the steady-state secular energy deviates from equipartition and becomes much higher than $W_n$. As $q$ increases, this deviation becomes significant more quickly. 

Second, the solution to Eq. \ref{eq:energy_relaxation_eq} is linear combination of three fundamental relaxation processes, whose rates are determined by the three eigenvalues of $\mathbf{M}$. The asymptotic behavior of the energy evolution is governed by the slowest relaxation rate, $\Gamma \lambda$, where $\lambda$, the smallest eigenvalue of $\mathbf{M}$, is
\begin{equation}
  \label{eq:lambda}
  \lambda = \frac{\tilde{m}}{(1+\tilde{m})^2}\left( 1-\frac{\tilde{m}}{\tilde{m}_c}\right)
\end{equation}
and $\tilde{m}_c$ is the critical mass ratio given in terms of trap parameters as, 
\begin{equation}
  \label{eq:m_c_3d}
  \tilde{m}_c = \frac{3(4-\alpha-4\epsilon+\sqrt{\alpha^2+8\alpha(1+\epsilon)+16\epsilon^2})}{4(2\epsilon +\alpha -1)}
\end{equation}

The eigenvalues of $\mathbf{M}$ are shown in Fig. \ref{fig:ke_comparison}(b) and are compared to the asymptotic relaxation rates observed in the simulation. For $\tilde{m}\ll \tilde{m}_c$, the cooling rate from Eq. \ref{eq:lambda} is similar to the traditional sympathetic cooling result up to a numerical factor \cite{DeCarvalho1999}. In this regime, the initial positive slope of $\lambda$ results from enhanced energy transfer efficiency through collisions with neutral atoms of similar mass. However, the additional factor $1-\frac{\tilde{m}}{\tilde{m}_c}$ causes $\lambda$ to reach a maximum and decrease to negative values once $\tilde{m}$ exceeds $\tilde{m}_c$. At this point, it is observed in the simulation that oscillation amplitude of the ion grows with collisions, until the ion becomes too energetic to be trapped, regardless of the energy of the buffer gas.

The transition from sympathetic cooling to heating by a buffer gas is thus defined by $\tilde{m} = \tilde{m}_c$ and is shown in Fig. \ref{fig:ke_comparison}(c) as a function of $q$ along with the results of the molecular dynamics simulations and previous results from other models of the process~\cite{Major1968, Moriwaki1992, DeVoe2009}. Taken together the results of Figs.~\ref{fig:ke_comparison}(a)-\ref{fig:ke_comparison}(c), make the case for using as small a buffer gas mass and as low $q$ as possible, if significant sympathetic cooling is desired.

\textit{Non-Maxwellian statistics in an ion trap} --
As originally observed in the seminal work of DeVoe~\cite{DeVoe2009}, the peculiarity of sympathetic cooling in ion trap is also manifested in the steady-state energy distribution of the ion, which features a heavy power-law tail due to the random amplifications of the ion energy by collisions. To gain a quantitative understanding of how this distribution arises, consider a simplified model, in which the motion of the ion and neutral atom' are restricted to one dimension, and $\mathcal{R}=-1$ in Eq. \ref{eq:post_collision_velocity}. In $(A, B)$ space, collisions result in a random walk given by 
\begin{equation}
  \label{eq:ab_jump}
  \begin{bmatrix}
    A_{N+1} \\
    B_{N+1}
  \end{bmatrix} = 
  \left(\mathbf{I} + \frac{\zeta}{w_0}
    \begin{bmatrix}
      s\dot{c} & s\dot{s} \\
      -c\dot{c}  &  c\dot{s}
    \end{bmatrix}_{\tau_N} \right)
  \begin{bmatrix}
    A_{N} \\
    B_{N}
  \end{bmatrix}  +  \frac{\zeta v_n}{w_0}
  \begin{bmatrix}
    s \\
    c 
  \end{bmatrix}_{\tau_N}
\end{equation}
where $\zeta=\frac{2\tilde{m}}{1+\tilde{m}}$, $[A_{N+1}, B_{N+1}]^\mathrm{T}$ are the coordinates after the $N$-th collision which occurs at $\tau=\tau_N$ ($N=1, 2, \cdots, \infty$). The $\tau_N$  constitute an array of Poissonian variables, with average interval equal to $\Gamma^{-1}$. As can be seen from Eq.~\ref{eq:ab_jump}, the random walk in $(A,B)$ space has both additive and multiplicative terms. As is well known in finance \cite{Sornette2001}, the multiplicative terms in the random walk give rise to the power law distribution as follows.

\begin{figure}[!]
  \centering
  \includegraphics[width=\columnwidth]{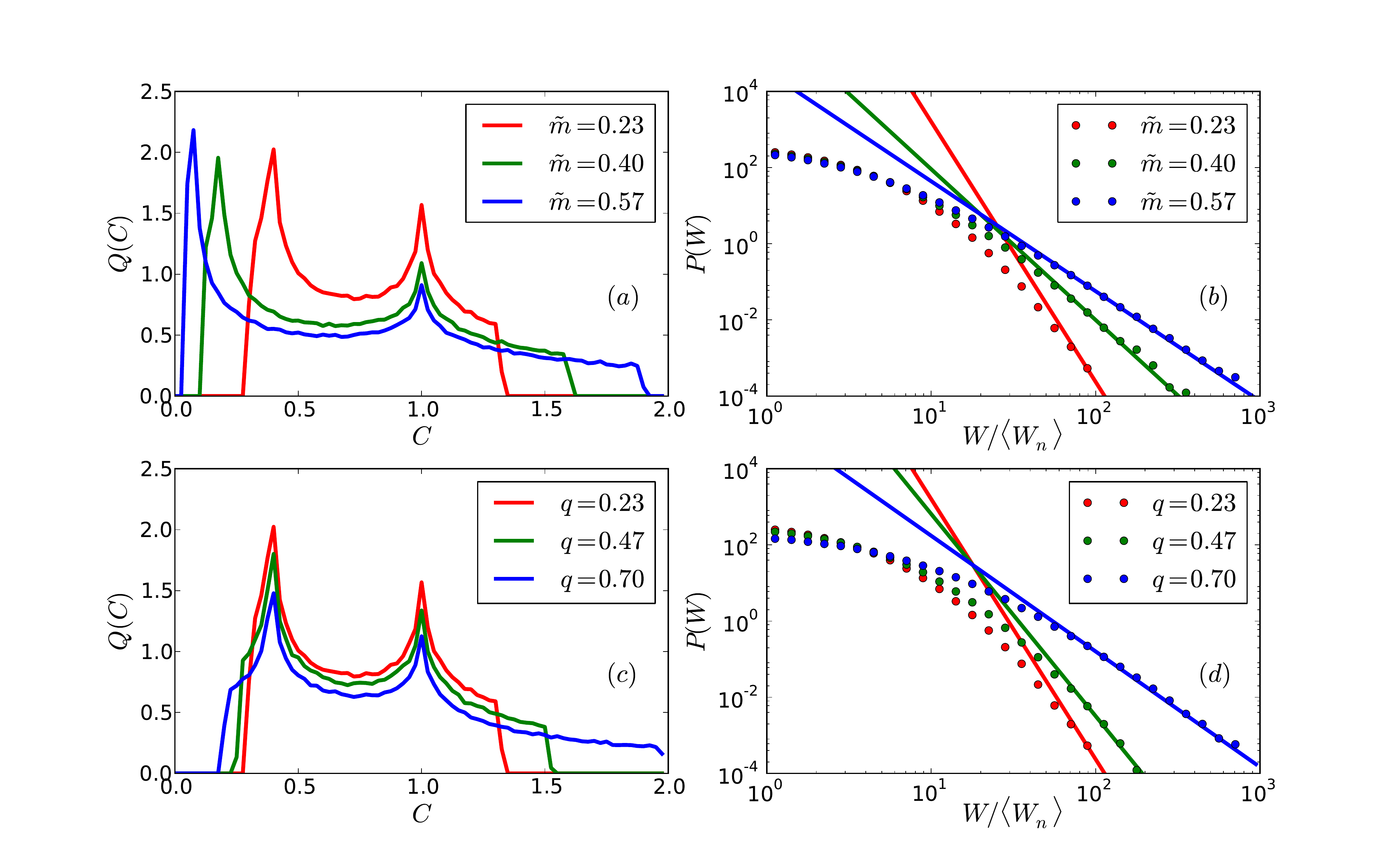}
  \caption{Probability density of the multiplicative noise $Q(C)$ and corresponding ion's energy $P(W)$ for 1D model from simulations for fixed $q = 0.23$ (lines in panel $a$ and $b$), and fixed $\tilde{m} = 0.23$ (dots in panel $c$ and $d$). The tail of $P(W)$ is fitted to the power-law form of $W^{-(\nu+1)}$ (solid line in panel $c$ and $d$), where $\nu$ is given by Eq. \ref{eq:nu}.}
  \label{fig:power_law}
\end{figure}

A recurrence relation for $W_N$ can be derived from Eq. \ref{eq:w_j}, and if only the distribution of high energy ions, i.e. $W_N \gg W_n$ is considered, this relation reduces to\begin{equation}
  \label{eq:Wn_recurrence}
  W_{N+1} = C W_N,
\end{equation}
where the multiplicative coefficient $C$ is given by,
\begin{equation}
  \begin{aligned}
    \label{eq:C}
  C(\tau_N, & \theta_N) = \cos^2\theta_N\left(\left(1+\zeta\frac{s\dot{c}}{w_0}\right)^2+\zeta^2\frac{c^2 \dot{c}^2}{w_0^2} \right)_{\tau_N}\\
  +& \sin^2\theta_N\left(\left(1-\zeta\frac{c\dot{s}}{w_0}\right)^2+\zeta^2\frac{s^2\dot{s}^2}{w_0^2} \right)_{\tau_N}\\
  +& 2\sin\theta_N\cos\theta_N\left(\zeta\frac{s\dot{s}-c\dot{c}}{w_0}+\zeta^2\frac{\dot{c}\dot{s}(c^2+s^2)}{w_0^2}\right)_{\tau_N}
  \end{aligned}
\end{equation}
and $\theta_N = \arctan(B_N/A_N)$. Because $W$ only depends on $A^2+B^2$, it is expected that as $N\rightarrow \infty$, $\theta_N$ becomes uniformly distributed in the range of $[0, 2\pi]$ and uncorrelated with $\tau_N$. $Q(C)$, the probability density of $C$, is calculated from Eq. \ref{eq:C} and exhibits random amplification of the ion energy, i.e. $C>1$, as shown in Fig. \ref{fig:power_law} panels $(a)$ and $(c)$ for different values of $\tilde{m}$ and $q$.

Due to this random amplification, $W$ develops a power-law tail in its probability density at steady state, i.e.  $P(W) \propto W^{-(\nu+1)}$~\cite{Takayasu1997}. Self-consistency requires that $P(W_{N+1})$, is equal to the product of $P(W_N)$ and $Q(C)$, under the constraint of Eq. \ref{eq:Wn_recurrence}, namely,
\begin{equation}
  \label{eq:nu_proof}
  \begin{aligned}
  P(W_{N+1}) & = \iint Q(C) P(W_N) \delta(W_{N+1}-CW_N)\;\mathrm{d}C\;\mathrm{d}W_N \\
  & = \int Q(C) P\left(\frac{W_{N+1}}{C}\right) \frac{1}{C}\;  \mathrm{d}C.
\end{aligned}
\end{equation}
Assuming $P(W) \propto W^{-(\nu+1)}$ then the power $\nu$ must satisfy
\begin{equation}
  \label{eq:nu}
  \langle C^{\nu} \rangle = 1.
\end{equation}

From this condition, $\nu$ can be found numerically and Fig.~\ref{fig:power_law}, panels $(b)$ and $(d)$, compare the prediction to the energy distribution extracted from a molecular dynamics simulation, which subjects the ion to $10^6$ trials, in each of which the ion undergoes $10^4$ collisions, for each $\tilde{m}$ and $q$ parameter. As $\tilde{m}$ and $q$ increase random amplification becomes more likely, causing the energy distribution to become more non-Maxwellian. In comparison, there is no such random amplification from collisions in a static trap (see Appendix D for details).

By considering the value of $\nu$ as $\tilde{m}\rightarrow0$ and $\tilde{m}\rightarrow\infty$, we find that the power can be approximated as $\nu_{\text{1D}} \approx 1.67/\tilde{m} - 0.67$ in 1D (see Appendix E). To extend the above discussion to a full 3D model, $C$ necessarily becomes a $3\times 3$ stochastic matrix, and the theory of stochastic matrix products \cite{Kesten1973}, which is beyond the current scope, must be considered. Nonetheless, one expects $\zeta_\text{3D} \approx \frac{1}{2}\zeta_{\text{1D}}$ because in 3D $\mathcal{R}$ average to zero, thus $\nu_{\text{3D}} \approx 2\nu_{\text{1D}}$, which agrees reasonably well with the empirically extracted power law of DeVoe, $\nu_{\text{emp}} \approx 4/\tilde{m}-1$. 
  
In summary, we have developed an analytical model that accurately predicts the steady state value and dynamics of the kinetic energy of a singe ion immersed in a neutral buffer gas. The transition from sympathetic cooling to heating, and its dependence on trap parameters and masses of the particles have also been explained. Finally, we have confirmed that the recent observation of non-Maxwellian statistics~\cite{DeVoe2009} for a trapped ion can be attributed to random heating collisions and provided a means to approximate the expected power law of the energy distribution. Taken together, these results solve the longstanding issues and questions that have existed since Dehmelt first considered this problem over forty years ago. We believe that these results will be critical for the design and interpretation of experiments in the rapidly growing field of hybrid atom-ion physics~\cite{Grier2009, Zipkes2010a, Zipkes2010b, Hall2011, Rellergert2011, Rellergert2013, Sullivan2012, Schmid2010a, Ratschbacher2012, Ratschbacher2013}.

\bibliography{reference_abbr}

\appendix
\renewcommand{\theequation}{A.\arabic{equation}}
\setcounter{equation}{0}

\section{Appendix A: Comparison of a quantum scattering calculation and $\frac{d\sigma_L}{d\Omega}$}
\begin{figure}[!]
  \centering
  \includegraphics[width=\columnwidth]{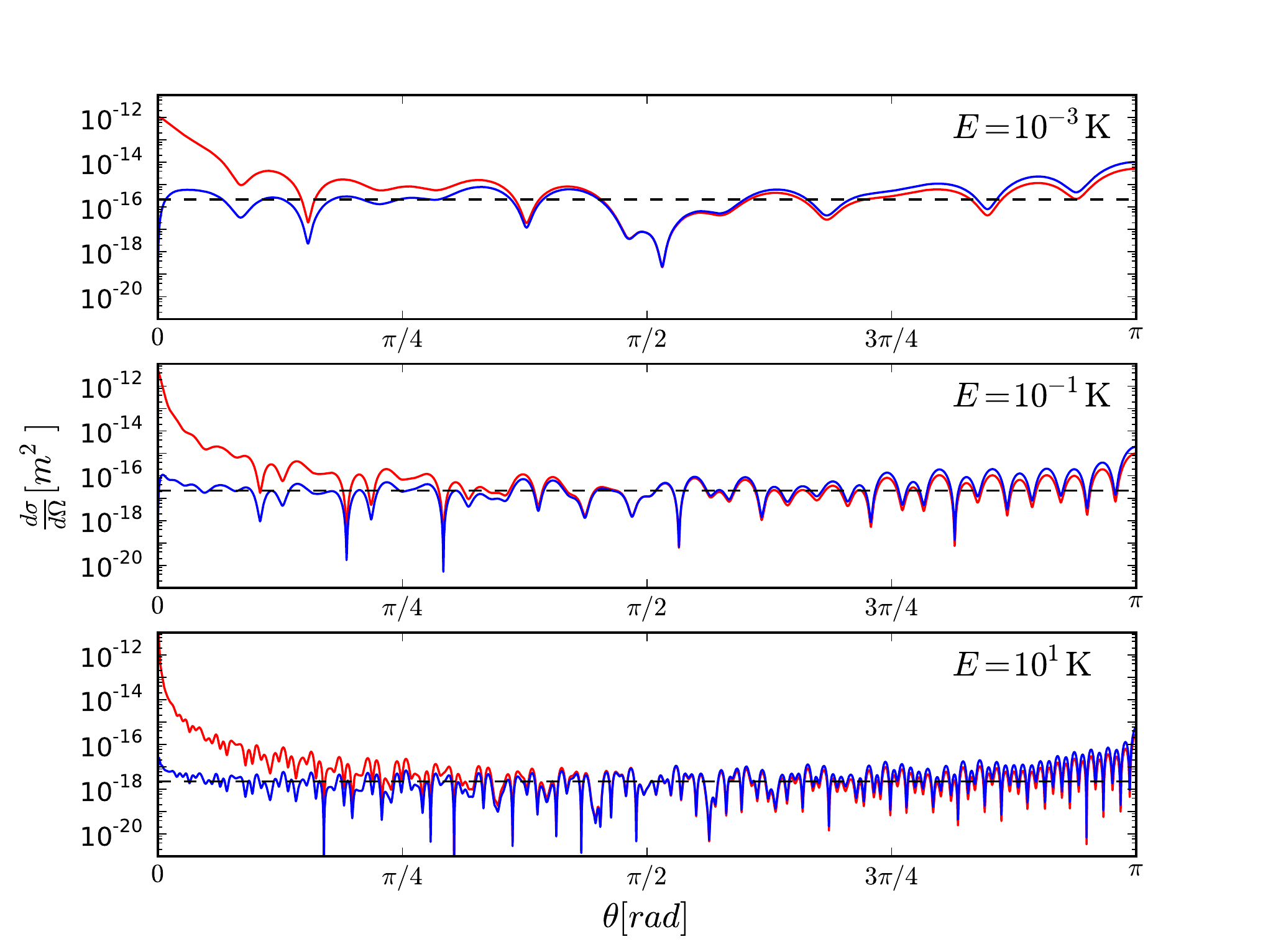}
  \caption{The elastic (red solid line), diffusion (blue solid line) and isotropic Langevin cross-section (black dashed line) for three different collision energy for the Yb$^{+}$ + Ca system.} 
  \label{fig:cross_section}
\end{figure}

In this section, we perform a quantum scattering calculation of the differential cross-section for the Yb$^+$ + Ca system, as the necessary interaction potential was available to us \cite{Rellergert2011}, and compare the results to an isotropic Langevin differential cross-section. 

Given the spherical symmetry of the atom-ion interaction potential, the differential cross-section can be calculated from \cite{Friedrich2005}
\begin{equation}
  \label{eq:diff_sigma_el}
  \frac{d\sigma_{el}}{d\Omega}(\theta, E) = \left| \sum_{\ell} (2\ell+1) P_\ell(\cos\theta) \frac{e^{2\imath\eta_\ell}-1}{2\imath k} \right|^2
\end{equation}
where $E$ is the collision energy and $\eta_\ell$ is the phase shift of the $\ell$-th partial wave induced by the interaction potential ~\cite{Johnson1999}. For a specific atom-ion combination, and thus for a specific interaction potential, it is straightforward to calculate this differential cross-section numerically.

The results are shown in Fig. \ref{fig:cross_section} for the Yb$^+$ + Ca system at three different energies and are expected to be similar for other atom-ion combinations~\cite{Zhang2009}. 

\begin{figure}[t!]
  \centering
  \includegraphics[width=\columnwidth]{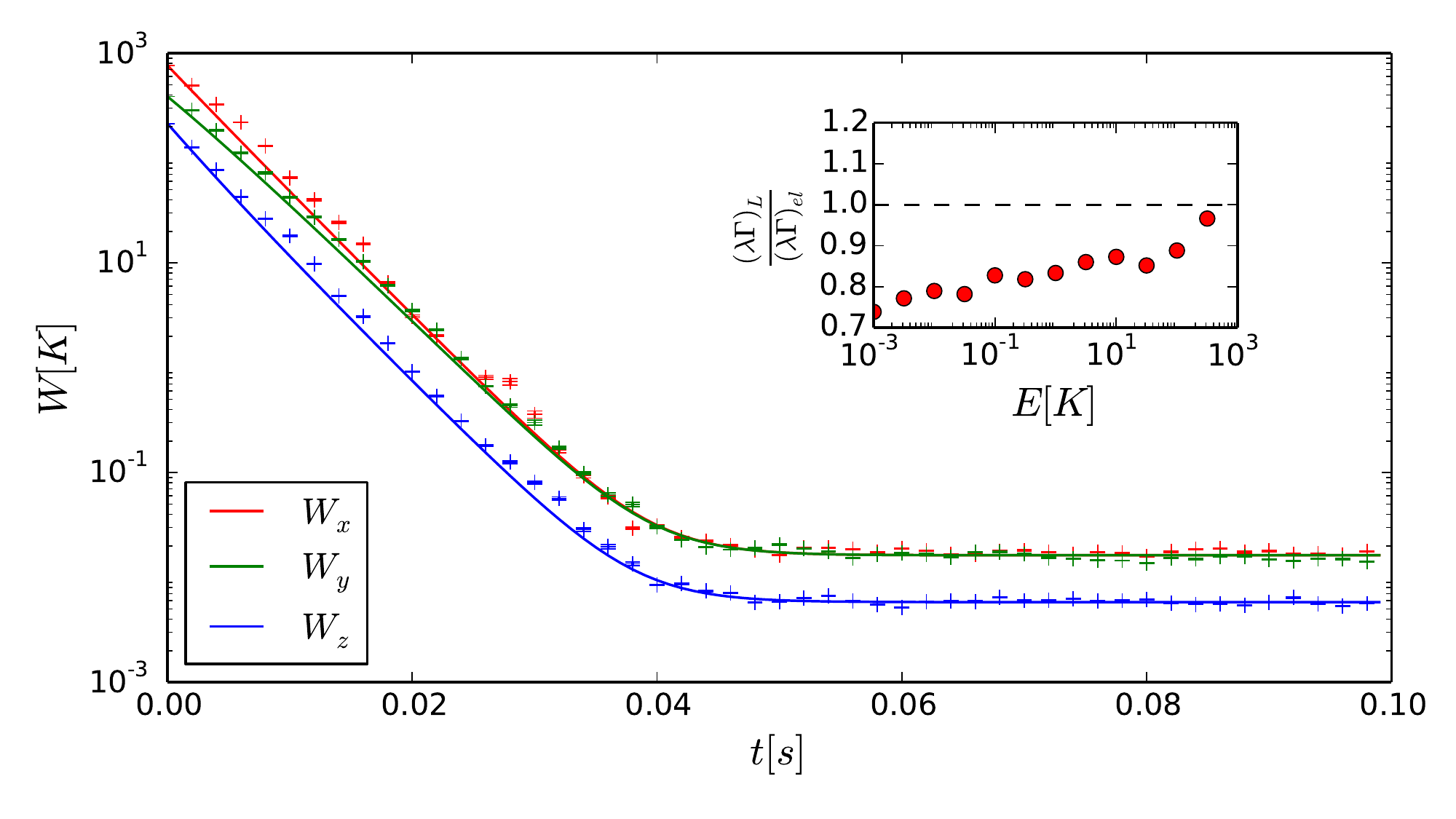}
  \caption{Simulation of the kinetic energy of a Yb$^+$ ion being sympathetically cooled by Ca atom ($T = 5\;\text{mK}$, $\rho = 8\times10^{11}\; \text{cm}^{-3}$) using $\sigma_{el}$, compared to the prediction from Eq. \ref{eq:energy_relaxation_eq} using $\sigma_{L}$ (lines). Inset: the ratio between aymptotic relaxation rates calculated using $\sigma_{L}$, and $\sigma_{el}$. To save simulation time, neutral atom's density $\rho = 8\times 10^{11}\mathrm{cm}^{-3}$.}
  \label{fig:rate_comparison}
\end{figure}

As can be seen in Fig.~\ref{fig:cross_section} the differential cross-section exhibits a large forward scattering peak at all energy scales. Thus, the majority of atom-ion collisions lead to only slightly deflected trajectories, resulting in a very small change in $\mathbf{W}$. Therefore, as originally argued by Dalgarno and co-workers~\cite{Dalgarno1958},  to prevent an overestimate of the energy redistribution due to collisions the momentum transfer (diffusion) differential cross-section, i.e. $\frac{d\sigma_d}{d\Omega}=\frac{d\sigma_{el}}{d\Omega}(1-\cos\theta)$, also shown in Fig.~\ref{fig:cross_section}, should be used to calculate the total atom-ion collision rate. Fortuitously, the diffusion differential cross-section is approximately isotropic in scattering angle, especially after thermal averaging, and agrees quite well with the simple Langevin cross-section~\cite{Langevin1905} $  \sigma_d \approx \sigma_L = \pi\sqrt{\frac{2C_4}{E}}$, as seen in Fig.~\ref{fig:cross_section}. Therefore, we replace the cross-section by an isotropic profile which integrates to $\sigma_L$. Under this approximation, the average collision rate $\Gamma = 2\pi\rho\sqrt{\frac{C_4}{\mu}}$ becomes energy independent and the calculation of $\mathbf{M}$ is greatly simplified. As demonstrated in Fig. \ref{fig:rate_comparison}, the validity of this approximation is confirmed by comparison to a detailed molecular dynamics simulation, which uses the full quantum differential cross-section. The resulting error in the relaxation rate is smaller than 25\% for collision energies down to 1~mK.

\renewcommand{\theequation}{B.\arabic{equation}}
\setcounter{equation}{0}

\section{Appendix B: Determination of M}
In this section, we provide more explicit details of the derivation of the relaxation matrix, $\mathbf{M}$.

First we rewrite, Eq.~\ref{eq:w_j} in terms of the instantaneous coordinates for the $x$ direction,
  \begin{align}
  \label{eq:w_xv}
    W_x = \frac{m\overline{\dot{c}_x^2}}{2w_{0,x}^2}\Big((\dot{c}_x^2+\dot{s}_x^2)x^2 +(c_x^2+s_x^2)v_x^2 \\ 
     - 2(c_x\dot{c}_x+s_x\dot{s}_x)xv_x \Big),
   \end{align}
where $w_{0,x} = c_x\dot{s}_x - s_x\dot{c}_x$ is the Wronskian and constant in time.  The change in average energy with a collision is then
 \begin{equation}
  \begin{aligned}
    \label{eq:delta_wx_2}
     W_x' -  W_x & =   - \frac{m\overline{\dot{c}_x^2}}{w_{0, x}^2}(c_x\dot{c}_x+s_x\dot{s}_x)(x( v_x' -v_x))\\
     & \;\;\;\; + \frac{m \overline{\dot{c}_x^2}}{2w_{0x}^2} (c_x^2+s_x^2)( v_x'^2-v_x^2) \\
     &\equiv \Delta W_{x,1} +  \Delta W_{x,2} . 
  \end{aligned}
\end{equation}

For $\Delta W_{x,1}$, since both $\langle \mathbf{v_n}\rangle$ and $\langle \mathcal{R}(\mathbf{v}-\mathbf{v_n}) \rangle $ vanish, $\langle v_{x}' \rangle = \frac{1}{1+\tilde{m}}v_x$. Therefore, using Eq. \ref{eq:ab_to_xv} and \ref{eq:w_j} we obtain,
\begin{equation}
  \label{eq:delta_wx_1}
  \begin{aligned}
  \langle \Delta W_{x,1} \rangle & = \frac{\tilde{m}}{1+\tilde{m}}\frac{m\overline{\dot{c}_x^2}}{w_{0,x}^2}\overline{(c_x\dot{c}_x+s_x\dot{s}_x)(x v_x)} \\
  & = \frac{\tilde{m}}{1+\tilde{m}} \frac{\overline{(c_x\dot{c}_x+s_x\dot{s}_x)^2} }{w_{0,x}^2} \frac{m}{2}\overline{\dot{c}_x^2}(a_x^2+b_x^2)\\
& = \frac{\tilde{m}}{1+\tilde{m}}\epsilon_x \langle W_x \rangle, 
  \end{aligned}
\end{equation}
where $ \epsilon_x  = \frac{\overline{(c_x\dot{c}_x+s_x\dot{s}_x)^2}}{w_{0,x}^2}$.

To evaluate $\Delta W_{x,2}$, $\mathbf{v_n}$, $\mathbf{v}$ and $\mathcal{R}$ are uncorrelated, the average value of cross-correlation terms between them vanish. Furthermore, since $\mathcal{R}$ is a random rotation, $\langle[\mathcal{R}(\mathbf{v}-\mathbf{v_n})]_x^2\rangle = \frac{1}{3}\langle(\mathbf{v}-\mathbf{v_n})^2\rangle$. Rearranging terms we obtain, 
\begin{align}
  \label{eq:delta_v_sq}
  \langle v_x'^2 \rangle - v_x^2 = \frac{\tilde{m}^2}{(1+\tilde{m})^2}\Big( \left(-\frac{2}{3}-\frac{2}{\tilde{m}}\right)v_x^2 + \frac{1}{3}v_y^2 \nonumber \\ 
   + \frac{1}{3}v_z^2 + 2\sigma_{v_n}^2 \Big) 
\end{align}
where $\sigma_{vn}^2 = 2\langle W_n\rangle/m_n$ is the thermal width of neutral atom velocity distribution. Thus, we find
%\begin{widetext}
\begin{equation}
  \begin{aligned}
  \label{eq:delta_wx_2}
  \langle \Delta W_{x,2} \rangle = \frac{\tilde{m}}{(1+\tilde{m})^2}\Big( \Big(-\frac{2\tilde{m}+2}{3}\Big)(1+\epsilon_x)\langle W_x \rangle + \\\frac{\tilde{m}\alpha_x}{6} \langle W_y \rangle +\frac{\tilde{m}\alpha_x}{6}\langle W_z \rangle + \alpha_x \langle W_n \rangle \Big), \\
\end{aligned}
\end{equation}
%\end{widetext}
where $\alpha_x  = \frac{\overline{(c_x^2+s_x^2)}\cdot\overline{(\dot{c}_x^2+\dot{s}_x^2)}}{w_{0,x}^2}$.

Combining the results of $\Delta W_{x,1}$ and $\Delta W_{x,2}$, and the results for the $y$ and $z$ directions, finally we have 
\begin{align}
    \langle\mathbf{W'}\rangle - \langle\mathbf{W}\rangle & = -\mathbf{M}\langle\mathbf{W}\rangle + \mathbf{N} \nonumber \\
    & = -\mathbf{M}(\langle\mathbf{W}\rangle-\mathbf{W}_{st})
  \end{align}
where  
  \begin{equation}
    \label{eq:M}
    \mathbf{M} = -\frac{\tilde{m}^2}{(1+\tilde{m})^2}
    \begin{bmatrix}
      \frac{2\epsilon-1}{3}-\frac{1}{\tilde{m}} & \frac{\alpha}{6} & \frac{\alpha}{6} \\
      \frac{\alpha}{6}  &           \frac{2\epsilon-1}{3}-\frac{1}{\tilde{m}}  & \frac{\alpha}{6} \\
      \frac{1}{6} &     \frac{1}{6} &  -\frac{1}{3}-\frac{1}{\tilde{m}}      \\
    \end{bmatrix}
  \end{equation}
and
  \begin{equation}
    \label{eq:N}
    \mathbf{N} = \frac{\tilde{m}}{(1+\tilde{m})^2}
    \begin{bmatrix}
      \alpha \langle W_n \rangle \\
      \alpha \langle W_n \rangle \\
      \langle W_n \rangle
    \end{bmatrix}
  \end{equation}
And, the steady-state kinetic energy is given by,
\begin{align}
  \label{eq:W_st}
  \mathbf{W}_{st} & = -\mathbf{M}^{-1}\mathbf{N} \nonumber \\ & =
\left(
  \mathbf{I} - \tilde{m}
  \begin{bmatrix}
    \frac{2\epsilon-1}{3} & \frac{\alpha}{6} & \frac{\alpha}{6} \\
    \frac{\alpha}{6} & \frac{2\epsilon-1}{3} & \frac{\alpha}{6} \\
    \frac{1}{6} & \frac{1}{6} &-\frac{1}{3}  \\
  \end{bmatrix}\right)^{-1}
  \begin{bmatrix}
      \alpha \langle W_n \rangle \\
      \alpha \langle W_n \rangle \\
      \langle W_n \rangle \\
  \end{bmatrix} 
\end{align}
where $\alpha \equiv \alpha_x = \alpha_y$ and $\epsilon \equiv \epsilon_x = \epsilon_y$. Because in the $z$ direction the trapping field is time-independent, $\alpha_z=1$ and $\epsilon_z=0$.
For low values of $q$ and $a$, the numerical values of $\alpha$ and $\epsilon$ are approximated by \cite{Chen2013},
\begin{align}
  \alpha \approx 2+2q^{2.24}\\
  \epsilon \approx 1+2.4q^{2.4}
\end{align}

\renewcommand{\theequation}{C.\arabic{equation}}
\setcounter{equation}{0}

\section{Appendix C: Procedures of Numerical Simulation}
We perform two types of Monte Carlo simulations to verify the analytical theory. Their simulation details are described below respectively. Type I simulations were initially carried out to verify that approximation of the differential scattering cross-section by an isotropic Langevin cross-section was valid (Fig. \ref{fig:rate_comparison}). Following the verification of the approximation, Type II simulations were used to make the simulations more computational efficient and resulted in the data for Figs. \ref{fig:ke_comparison}(a)-(c).

\subsection*{Type I}
In Type I simulations, the ion trajectory is found numerically by integrating the equations of motion with fixed time step $\Delta t$ using a custom modified version of the ProtoMol software \cite{Matthey2004}, where $\Delta t$ is chosen to be much smaller than the rf period $\Omega^{-1}$. The differential elastic collision cross-section $\frac{d\sigma_{el}}{d\theta}$ obtained from a quantum scattering calculation is used in every collision. The simulation consists of following four steps:

  \begin{enumerate}
    \renewcommand{\theenumi}{S\arabic{enumi}}
  \item A single ion is initialized at the origin with zero velocity, i.e. $\mathbf{r}_0=\mathbf{0}$, and $\mathbf{v}_0=\mathbf{0}$. The simulation step index $N$ is set to 0.

  \item The position and velocity of the ion, $\mathbf{r}_{N+1}$ and $\mathbf{v}_{N+1}$, at the next step $N+1$ are calculated by leapfrog integration of the equations of motion.

  \item To determine if a collision should happen during $\Delta t$, an atom  is generated with velocity $\mathbf{v_n}$ sampled from thermal distribution characterized by $W_n$. The associated collision rate $\Gamma$ is given by $\rho\sigma_{el}|\mathbf{v_{rel}}|$, where $\rho$ is the density of ultracold atoms,  $\mathbf{v_{rel}}$ is the relative velocity, and $\sigma_{el}$ depends implicitly on the collision energy $\frac{\mu}{2}\mathbf{v_{rel}}^2$ in the center-of-mass frame. A collision happens during $\Delta t$ if $1-\exp(-\Gamma \Delta t) < d$, where $d$ is the value of a random number chosen from a uniform distribution over$[0, 1]$. If this condition is met the simulation then proceeds to S4, otherwise it returns to S2.

  \item The velocity of the ion after the collision is updated according to Eq. \ref{eq:post_collision_velocity}. The rotation matrix $\mathcal{R}$ is specified by polar angle $\theta$ and azimuthal angle $\phi$, defined with respect to $\mathbf{v_{rel}}$. $\theta$ is sampled from the probability distribution function $\frac{d\sigma_{el}}{d\theta}\sin\theta$ defined on $[0, \pi]$, and $\phi$ is sampled from uniform distribution on $[0, 2\pi]$. The simulation then loops back to S2 until the prescribed number of collisions have been reached.
  \end{enumerate}

\subsection*{Type II}
  In Type II simulations, the isotropic Langevin differential cross-section $\frac{d\sigma_L}{d\Omega}$ is used to calculate scattering process. The collision rate $\Gamma$ thus does not depend on collision energy, allowing for a much faster integration method based on a transfer matrix similar to Ref.~\cite{DeVoe2009}. The simulation consists of the following four steps,
  \begin{enumerate}
    \renewcommand{\theenumi}{S\arabic{enumi}}
  \item A single ion is initialized at the origin with zero velocity, i.e. $\mathbf{r}_0=\mathbf{0}$, and $\mathbf{v}_0=\mathbf{0}$, and, a series of collision times $\tau_j~(j=1, 2, 3, \cdots)$ are pre-determined, which follow a Poissonian distribution with average interval equal to $\Gamma^{-1}$.

  \item The new coordinate of the ion  $\mathbf{P}_{i+1} = [x_i, v_{x,i}, y_{i+1}, v_{y,i+1}, z_{i+1}, v_{z,i+1}]^\mathrm{T}$ at $\tau=\tau_{i+1}$ is obtained by multiplication of $\mathbf{P}_i$ by the transfer matrix $\mathbf{T}(\tau_{i+1}, \tau_i)$~\cite{DeVoe2009}. $\mathbf{T}$ consists of three $2\times2$ submatrices,
    \begin{equation}
      \label{eq:big_M} \mathbf{T} = 
      \begin{bmatrix}
        \mathbf{T}_x & \mathbf{0} & \mathbf{0} \\
        \mathbf{0} & \mathbf{T}_y & \mathbf{0} \\
        \mathbf{0} & \mathbf{0}   & \mathbf{T}_z 
      \end{bmatrix}
    \end{equation}
where each submatrix $\mathbf{T}_j$ ($j = x, y, z$) is given by 
\begin{widetext}

    \begin{equation}
      \mathbf{T}_j(\tau_2, \tau_1) = \frac{1}{w_{0,j}}
      \begin{bmatrix}
        \label{eq:transfer_matrix}
        c_j(\tau_2)\dot{s}_j(\tau_1) - s_j(\tau_2)\dot{c}_j(\tau_1) & -c_j(\tau_2) s_j(\tau_1) + s_j(\tau_2)c_j(\tau_1)\\
         \dot{c}_j(\tau_2) \dot{s}_j(\tau_1) - \dot{s}_j(\tau_2) \dot{c}_j(\tau_1) & -\dot{c}_j(\tau_2) s_j(\tau_1)  +\dot{s}_j(\tau_2) c_j(\tau_1) 
      \end{bmatrix}
    \end{equation}
\end{widetext}

  \item A collision then modifies the velocity of the ion according to Eq. \ref{eq:post_collision_velocity}, where $\mathcal{R}$ now represents a rotation with equal probability into a $4\pi$ solid angle. The simulation then loops back to S2, until the prescribed number of collisions has been reached.
  \end{enumerate}

\renewcommand{\theequation}{D.\arabic{equation}}
\setcounter{equation}{0}

\section{Appendix D: The lack of multiplicative noise in a static trap}
In sharp contrast to the ion trap case, a particle confined in a static potential $V(x) = \frac{m}{2}\omega^2 x^2$ and in contact with a reservoir at temperature $T$ would have the same thermal distribution, regardless of the reservoir particle's mass $m_n$, or the trapping frequency $\omega$. This is because for static traps the Mathieu functions $c(\tau)$ and $s(\tau)$ are replaced by $\cos(\omega \tau)$ and $\sin(\omega \tau)$, which simplifies Eq. \ref{eq:C} into
\begin{equation}
  \label{eq:C_for_static_trap}
  C = 1 - (2-\zeta)\zeta\sin^2(\omega\tau-\theta) 
\end{equation}
Since $0\le\zeta\le 2$, $C\le 1$, thus the energy of such system is never amplified. From a mathematical perspective, the solution for Eq. \ref{eq:nu} is $\nu \rightarrow \infty$, meaning the predicted energy distribution falls faster than any power-law tail of finite $\nu$, consistent with the thermal distribution $\exp(-E/k_BT)$.

\renewcommand{\theequation}{E.\arabic{equation}}
\setcounter{equation}{0}

\section{Appendix E: Determination of $\nu_{\text{1D}}$}
To determine $\nu_{\text{1D}}$, first consider the light buffer-gas mass limit \textit{i.e.} $\tilde{m}\rightarrow 0$, and $\zeta \approx 2\tilde{m}$. Ignoring $O(\zeta^2)$, $C$ in Eq. \ref{eq:C} is simplified to:
\begin{equation}
  C =  1 - \zeta + \zeta \delta 
\end{equation}
where $\delta = \left(\frac{c\dot{s}+s\dot{c}}{w_0}\right)_{\tau+\theta}$. 

\begin{figure}[t!]
  \centering
  \includegraphics[width=\columnwidth]{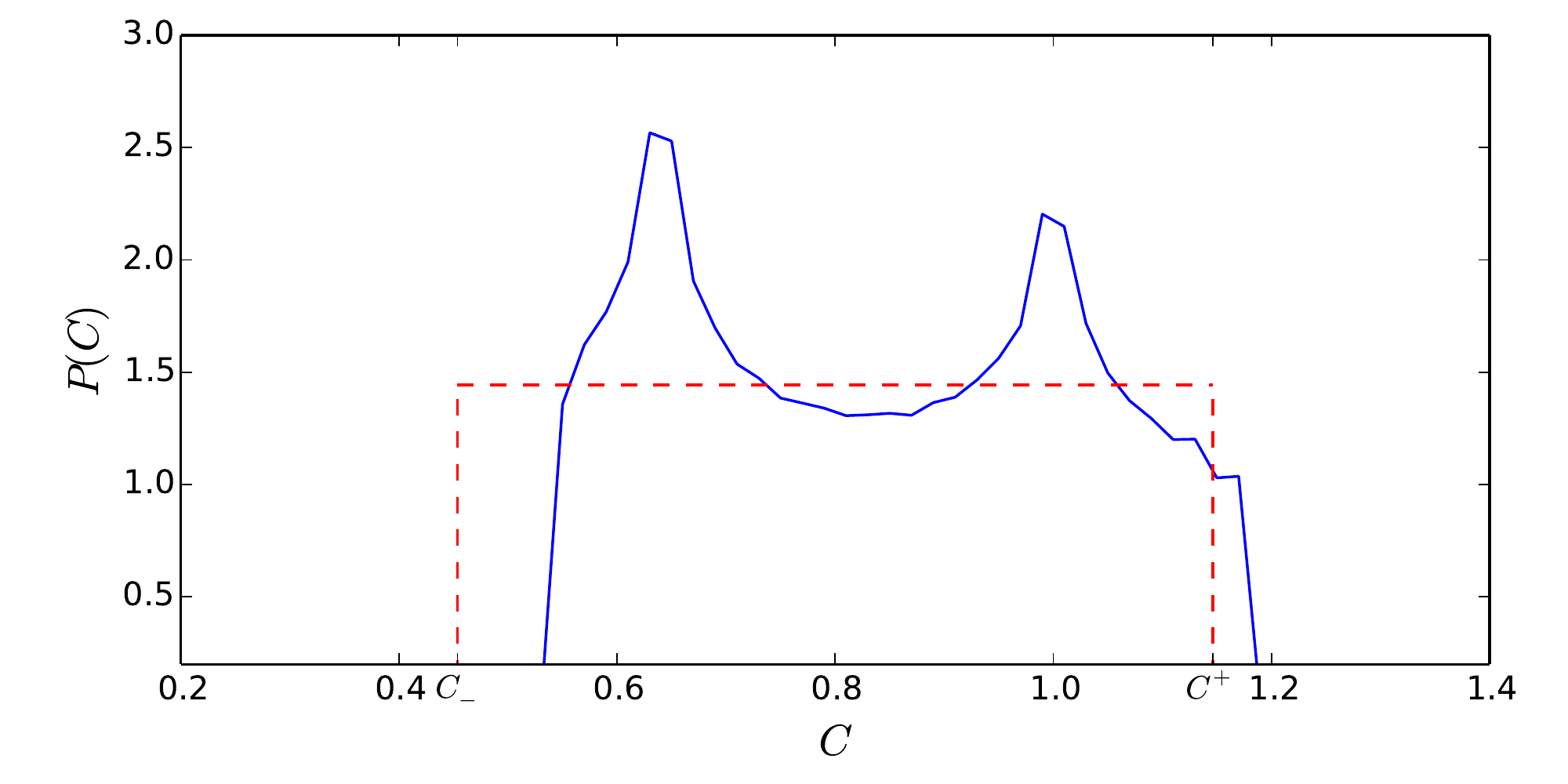}
  \caption{Exact value of $P(C)$ (blue solid line) sampled from Eq. \ref{eq:C} for $\zeta=0.2, q=0.1$, and the uniform approximation $\tilde{P}(C)$ (red dashed line).} 
  \label{fig:p_delta}
\end{figure}

The analytical form of $P(C)$ is difficult to calculate. Instead, we approximate it by a uniform distribution $\tilde{P}(C)$ in the range of $[C_{-}, C_{+}]$, which preserves the value of first and second moment of $C$, namely $\langle C \rangle_{P} = \langle C \rangle_{\tilde{P}}$, and $\langle C^2 \rangle_{P} = \langle C^2 \rangle_{\tilde{P}}$, where the subscript denotes the distribution for which the average value is calculated. With this requirement, $C_\pm$ is given by
\begin{equation}
  C_\pm = 1 - \zeta \pm \zeta \delta_m
\end{equation}
where $\delta_m = \sqrt{3 \langle\delta^2\rangle} \approx \sqrt{3}$  for $q < 0.4$. Thus,
\begin{equation}
  \tilde{P}(C) =
  \begin{cases}
    \frac{1}{2\zeta \delta_m} & \text{if } C \in [C_-, C_+] \\
    0 & \text{otherwise}
  \end{cases}
\end{equation}
An example of $\tilde{P}(C)$ is shown in Fig. \ref{fig:p_delta}.

With $\tilde{P}(C)$, we solve for $\nu$ with a straightforward calculation of $\langle C^\nu \rangle$,
\begin{equation}
  \langle C^\nu \rangle_{\tilde{P}} = \int^{C_+}_{C_-} \frac{C^\nu}{2\zeta \delta_m} \;\text{d}C = \frac{1}{2\zeta \delta_m} \frac{(1 + \zeta (\delta_m-1))^{\nu+1}}{\nu+1}  = 1
\end{equation}
Note $C_-^{\nu+1}$ vanishes because $C_- < 1$ and $\nu\gg 1$. Introducing $k = \zeta (\nu+1)$, we get
\begin{equation}
  \frac{1}{2\delta_m} \frac{(1 + \zeta (\delta_m-1))^{k/\zeta}}{k} \approx \frac{e^{(\delta_m-1)k}}{2\delta_m k} = 1
\end{equation}
with the value of $k \approx 3.35$ solved numerically. Thus we have the scaling relation for $\tilde{m}\rightarrow 0$,
\begin{equation}
  \label{eq:nuapprox}
  \nu_{\text{1D}} \approx \frac{3.35}{\zeta} \approx \frac{1.67}{\tilde{m}} 
\end{equation}

\begin{figure}[t!]
  \centering
  \resizebox{1\columnwidth}{!}{\input{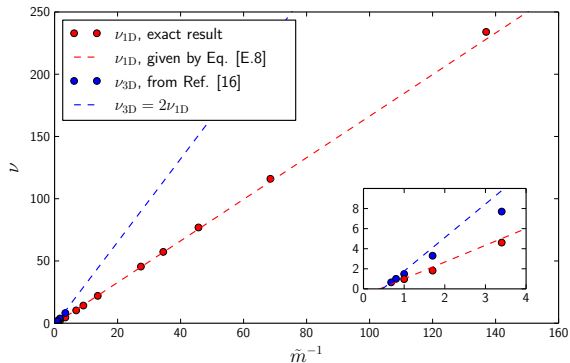}}
  \caption{Comparison of exact solution (red dots) of $\nu_{\text{1D}}$ with result calculated by Eq. \ref{eq:nuapproxbetter} (red dashed line). For reference, also shown are $\nu_{\text{3D}}$ from Ref. \cite{DeVoe2009} (blue dots) and an estimation of $\nu_{\text{3D}} = 2\nu_{\text{1D}}$ (blue dashed line). }
  \label{fig:nu_approx_better}
\end{figure}

Now consider the heavy buffer gas limit where $\tilde{m}\rightarrow \tilde{m}_c$. Clearly we must have, 
\begin{equation}
  \nu_{\text{1D}}(\tilde{m} = \tilde{m}_c)=1
  \label{eq:nu_at_mc}
\end{equation}
since when $\tilde{m} = \tilde{m}_c$, $\langle C \rangle = 1$ and the variances of $A$ and $B$ diverge \cite{Takayasu1997}.

Our previous approxmations break down because the $\zeta^2$ term cannot be ignored. Thus, we do not seek to carry out further analysis, but instead add an intercept to Eq. \ref{eq:nuapprox} such that the requirement Eq. \ref{eq:nu_at_mc} is met. For $q <0.4$ we find 
\begin{equation}
  \label{eq:nuapproxbetter}
  \nu_{\text{1D}} = \frac{1.67}{\tilde{m}} - 0.67,
\end{equation}
which agrees surprisingly well with the exact value of $\nu_{\text{1D}}$ (shown in Fig. \ref{fig:nu_approx_better}). 

\end{document}